# Temperature-dependent photoionization thresholds of alkali-metal nanoparticles reveal thermal expansion and the melting transition

*Abdelrahman O. Haridy,*[*] *Atef A. Sheekhoon,*[*] *Vitaly V. Kresin*

Department of Physics and Astronomy, University of Southern California,

Los Angeles, 90089-0484, USA

A precision measurement of the photoionization of pure sodium and potassium nanoparticles isolated in a beam enabled an accurate determination of their work functions as a function of temperature. In addition to resolving and quantifying the initial gradual decrease of the work function with temperature. which is associated with thermal expansion, the experiment revealed that the work function then undergoes a distinct drop both in magnitude and in slope that signifies the onset of nanoparticle melting. This establishes that a structural phase transition can be detected via a high-resolution measurement of the photoemission threshold. The melting temperature of nanoparticles with diameters of 7–9 nm is reduced by nearly 100 K relative to the bulk value. This suppression aligns with predictions from the Gibbs–Thomson equation which describes finite-size phase transitions.

[*]These authors have contributed equally

*Introduction.* Phase change behavior in nanoclusters and nanoparticles can deviate significantly from that of bulk materials. As the particle size decreases, the surface-to-volume ratio increases, leading to pronounced effects. One of them is the substantial reduction in the melting temperature which is largely due to the heightened role of surface curvature at the nanoscale. This phenomenon is commonly referred to as a manifestation of the "Gibbs-Thomson" (or "Gibbs-Kelvin") equation and is described in many publications (see, e.g., [1-6] and references therein).

For small particles on substrates and for microcrystals this behavior has been investigated using electron microscopy. The advantage of this technique is that structural changes can be imaged directly (although care has to be taken not to allow the electron beam itself to affect the structure). Alternatively, isolated nanoparticles and nanoclusters in beams have the benefits of being free from the influence of a substrate and of being amenable to mass spectrometry, but direct structural imaging of the melting of nanometer-scale particles in a beam is so far impractical. (The recently achieved coherent diffractive imaging of surface and bulk melting in much larger free nanoparticles [7-9] is likely to be extended to smaller system in the future.) Hence there is a need for additional techniques for the detection of phase transitions.

One important tool is in-flight calorimetry [10,11]. Heat capacities are deduced from the fragmentation patterns produced by laser heating of prethermalized clusters, and peaks in the heat capacity curves can represent the onset of melting. Another method is ion mobility measurements, where cluster ions drift through an inert gas under the influence of an electric field [12]. Melting alters their geometrical cross sections and consequently their drift velocities.

Here we describe a new approach which is based on measuring the temperature dependence of the ionization energies of metal nanoparticles. Materials expand as they are heated, and upon melting their densities may change abruptly. The ionization energy (and its bulk counterpart, the work function) depends on the electron density and therefore experiences a corresponding shift. Thus by exploiting the connection between electronic and structural degrees of freedom, the evolution of the photoionization spectrum can be used as a probe of the underlying structural dynamics.

The connection between the photoelectric effect and the melting transition has remained largely unexplored, despite having the potential to serve as a probe of short- and long-range order in materials [13]. This gap stems from the fact that the temperature dependence of work functions in general is weak [14] and has attracted only a relatively limited amount of attention [15-25]. We have developed a setup and an analysis procedure that are capable of highly precise determination of the ionization energies of temperature-controlled alkali metal nanoparticles in a beam [26]. A major benefit of the use of free nanoparticles is that their flight time through the evacuated apparatus is only a few milliseconds, therefore they avoid contamination which can easily skew the photoemission properties of highly reactive surfaces. The approach described below makes it possible to resolve shifts in electronic structure that accompany a structural phase transition, providing a distinct signature of thermal expansion and melting in nanoscale systems.

*Methods.* In the apparatus described in detail in [26], alkali-metal nanoparticles were generated by evaporating metal from a heated crucible and rapidly capturing and cooling the vapor within a flow of cold helium gas. This is often referred to as a gas aggregation source [27,28]. The nucleated nanoparticles were then transported by the gas through a thermalization tube where their temperature could be adjusted between 60 K and 400 K, and exited into vacuum as a directed beam. Since the helium pressure is relatively low (0.1 – 1 mbar) and the tube opening is relatively large (6 mm), the expansion of the gas-nanoparticle mixture into vacuum will not measurably change the temperature of the particles. This is supported, for example, by data on sodium cluster ions with ~$10^2$ atoms [29], where a flow tube with similar parameters was shown to reliably thermalize the clusters to the same temperature as a buffer-gas trap. The nanoparticles in the present experiment are an order of magnitude larger (see below) and therefore would be even less liable to experience a temperature change upon exit from the tube.

In this context, a remark also is in order regarding the effect of possible in-flight evaporative cooling of the nanoparticles, especially at the high-temperature range of the measurements. As is shown in the Appendix, for the sizes studied here this will not result in substantial in-flight cooling of the nanoparticles either.

After a free flight path, the nanoparticles entered the photoionization and detection chamber. Here they were ionized by an arc lamp through a monochromator, and the resulting positive ions were detected and counted. The yield $Y$ of photoions, equivalent to that of photoelectrons, was normalized to both photon and particle fluxes.

For nanoclusters and nanoparticles the near-threshold shape of the photoionization curve has been found (see [26,30] and references therein) to be very accurately fitted by the Fowler formula for the yield of photoelectrons from metal surfaces [31]:

$$\log\left(\frac{Y}{T^2}\right) = B + \log f\left(\frac{h\nu - I}{k_B T}\right) \tag{1}$$

Here $h\nu$ is the photon energy, $I$ is the ionization energy (or work function, in the bulk limit), $B$ a temperature-independent coefficient incorporating fundamental constants and instrumental parameters, and $f$ is an integral over a segment of the Fermi-Dirac distribution function.

For a quantitative analysis the nanoparticles' size distribution also must be taken into account. It was measured by a Wiley-McLaren time-of-flight mass spectrometer set up collinearly with the beam machine axis, using third harmonic pulses from a Nd-YAG laser. The size distribution for Na and K was well described by a lognormal function [32]. Under the given experimental conditions [26] the average size was found to be $N \approx 5000$ and the full-width-at-half-maximum was of the same magnitude, $\Delta N \simeq N$.

In view of the particle size dispersion and the fact that the average size fluctuated between runs (by approximately ±15%, depending on the thermalization tube temperatures and precise condensation source conditions), it is useful to represent the data by scaling it in a consistent way, since smaller nanoparticles have higher ionization energies,. We do this by plotting the ionization energy $I$ extrapolated to the bulk work function limit by means of the scaling relation

$$I = W + \alpha e^2 / R . \tag{2}$$

Here $R$ is the particle radius:

$$R = r_s a_0 N^{1/3}, \tag{3}$$

with $a_0$ the Bohr radius and $r_s$ the Wigner-Seitz electron density parameter [33] (3.9 for Na, 4.9 for K). We use the experimentally reported values of $\alpha \approx 0.39$ for Na [34] and $\alpha \approx 0.34\text{-}0.38$ for K [35,36]. These are close to $\alpha = 3/8$ predicted by an electrostatic image charge argument [37], although $\alpha = 1/2$ also has been suggested [38].

Thus for an individual measurement at a given nanoparticle temperature and a known size distribution, we use $I$ expressed in terms of $W$ and $R$ according to Eq. (2) in the Fowler equation (1), and convolute the latter [26] with the specific particle size distribution determined during the same measurement. The resulting value of $W$ fitted to the photoionization yield data can be employed on a single plot as a characteristic of the overall nanoparticle population. This is presented in the next section.

*Melting temperature of nanoparticles.* Fig. 1 presents temperature-dependent values of the work function $W$ for free Na and K nanoparticles, determined as described above. The key observation is that the data for each element exhibit two distinct regimes characterized by different slopes and separated by a discontinuity. It is rational to assign this feature to the nanoparticle melting transition. As described below, its parameters are consistent with this interpretation. To the best of our knowledge, this is the first detection of the melting transition in nanoparticles via its effect on the electronic work function.

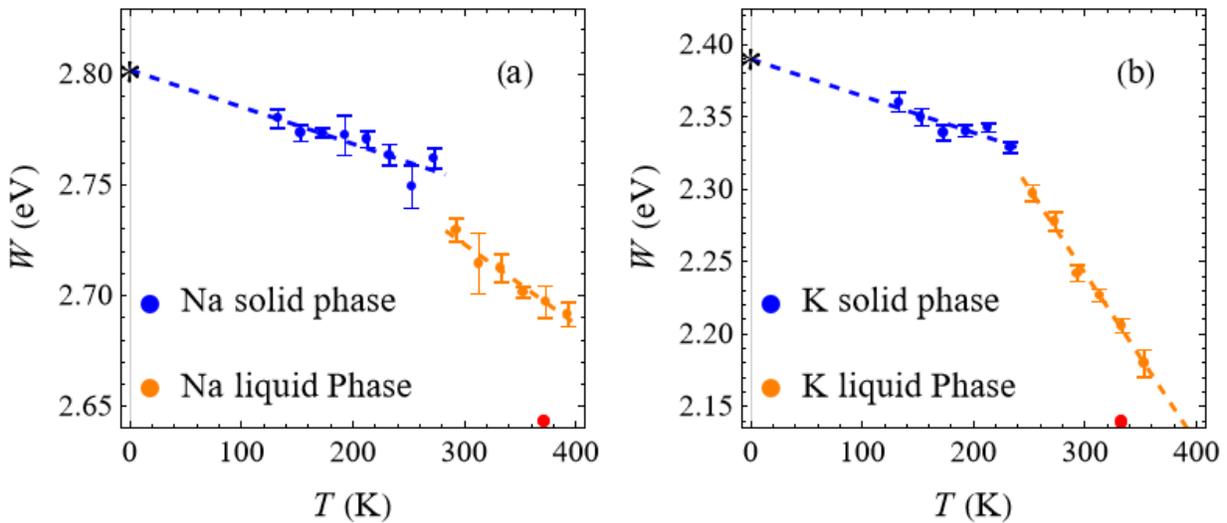

**Fig. 1** The experimentally determined work functions of (a) Na and (b) K as a function of temperature. The dashed straight lines are least-squares fits. The error bars show the calculated standard error after three individual ionization measurements for each temperature. The dots on the *x*-axis indicate the bulk melting point for each metal. The asterisks mark the linear extrapolation of the experimental data points to $T = 0$ K.

Importantly, the transitions are seen to occur at temperatures much lower than the bulk metal melting points marked on the axis of Fig. 1. As mentioned in the Introduction, this melting point depression is a well-established phenomenon in nanoscale systems, stemming from the enhanced contribution of surface energy.

In [39] the disappearance of geometric shell structure in $Na_N$ clusters with increasing temperature was observed and interpreted as evidence for size-dependent melting. For $N$ of comparable size this found to occur at 300 K, which is quite close to the location of the discontinuity in Fig. 1(a). This lends additional support to our assignment of the latter to the melting transition.

For spherical particles of radius $R$, the Gibbs–Thomson equation for the melting point reads

$$T_m(R) = T_m \left( 1 - \frac{c\sigma_{sl}}{H_f R} \right) \qquad (4)$$

Here $T_m$ is the melting temperature in the bulk limit, $\sigma_{sl}$ is the solid-liquid interfacial energy per unit area, and $H_f$ is the enthalpy of fusion (melting) per unit volume. In the thermodynamic phase equilibrium limit the prefactor is typically given as [40] $c = 2$, but for isolated nanoparticles in the gas phase the value $c = 3$ has been derived [32].

Experimental data on solid–liquid interfacial energies for alkali metals are scarce in general, primarily due to their high chemical reactivity, and are available only for bimaterial interfaces. Therefore we refer to the widely used estimate due to Turnbull [41] which relates the interfacial energy to the enthalpy of fusion. Expressed in terms of quantities per unit area and volume, it reads

$$\sigma_{sl} \approx \sigma_0 H_f d_m, \qquad (5)$$

where the atomic volume is defined as $d_m^3 = V_m / N_A$ ($V_m$ is the molar, or "gram atomic," volume and $N_A$ is the Avogadro number).

Empirically, the coefficient $\sigma_0$ is found to be $\approx 0.45$ for many metals [41]. Additional modifications and size corrections to the interfacial energy and enthalpy factors in Eq. (4) can be found in the literature [42-45]. Here we restrict ourselves to the compact expression (5). Inserting it into Eq. (4) results in the estimate

$$T_m(R) \approx T_m\left(1 - c\sigma_0 \frac{d_m}{R}\right) \approx T_m\left(1 - \frac{1.6c\sigma_0}{N^{1/3}}\right). \tag{6}$$

In the last term we used Eq. (3) together with the fact that for monovalent metals the atomic volume $d_m^3$ is also given by $\frac{4}{3}\pi(r_s a_0)^3$. Fig. 1 shows that the phase transitions for Na and K nanoparticles are observed respectively at approximately 280 K and 240 K, corresponding to $T_m(R)/T_m \approx 0.75$ and $\approx 0.7$. The rudimentary equation above correctly predicts that the fractional shift should be of similar magnitude for both materials. With $c = 3$ and $N \approx 5000$, it yields a shift within approximately a factor of two of the experimental value.

*Work function change upon melting.* The melting transition is typically accompanied by a drop in the mass density $\rho$ due to the less compact arrangement of atoms in the molten phase. This change will manifest as a discontinuity in the work function $W$. If we assume that the temperature slope of $W$ is primarily due to thermal expansion, we can estimate the discontinuity in $W$ due to the jump in density as follows:

$$\Delta W \approx \frac{dW}{d\rho}\Delta\rho = \frac{dW}{dT}\frac{dT}{d\rho}\Delta\rho = \frac{dW}{dT}\frac{1}{\beta}\frac{\Delta\rho}{\rho}, \tag{7}$$

where $\beta$ is the coefficient of volume thermal expansion. Table I compares the work function change seen in Fig. 1, with that calculated from Eq. (7) using reported melting-induced density changes of Na and K [46,47], the solid-phase slopes of $W(T)$ from Fig. 1 (also discussed in the next section), and the corresponding thermal expansion coefficients [48,49]. Considering the uncertainties in the input parameters, the agreement is reasonable.

*Temperature dependence of the work function.* Table II lists the values to which the Na and K work functions plotted in Fig. 1 extrapolate for $T = 0$ K. (These are 50-80 meV higher than an earlier determination [22] which was performed by a similar method but fewer data points and with less accurate temperature control and threshold fitting procedure.) These $T = 0$ work functions are slightly higher than the values quoted in literature reviews. However, the latter are room temperature values whereas Fig. 1 shows that the work functions decrease with temperature. Therefore the table also lists the result of extrapolating the solid-state phase lines in Fig.1 to room temperature (i.e., beyond the nanoparticles' melting points).

**Table I.** Magnitudes of melting-induced work function drops estimated from the data in Fig. 1 and from Eq. (7).

|   | Measured $|\Delta W|$ (eV) | Calculated $|\Delta W|$ (eV) |
|---|---|---|
| Na | 0.025-0.03 | 0.033 |
| K | 0.02-0.03 | 0.046 |

In point of fact, the variation of work functions with temperature is a topic dating back to studies of thermionic emission (see the references cited in the Introduction). Nonetheless, even at present there doesn't appear to exist a comprehensive microscopic theory of this behavior. As enumerated in the review [16], the dependence $W(T)$ in principle has contributions from static thermal expansion of the material (change in the bulk density and the size of the surface dipole layer), from changes in the dynamical electron-phonon interaction, and from changes in the electron effective mass and in the state occupation function. Of these, the first factor (density change with temperature) may have the most significance for the simple alkali metals.

A phenomenological approximation that has been found to provide a good degree of agreement with the magnitude and scaling of work functions data is based on the image charge approximation. Here $W$ is treated as the work required to remove an electron from a distance $d$ outside the planar surface to infinity, where $d$ is on the order of the screening length; see the discussion and references in [50]. Combining the density dependence of $d$ with the materials' thermal expansion coefficients, the latter authors then systematically calculated [21] the temperature variation of $W$ for polycrystalline metal surfaces, fitting it to a polynomial form $dW/dT = a + bT$.

Table II displays the slopes of the solid-phase experimental data from Fig. 1, as well as linear interpolation of the aforementioned model [21] values extended to the same temperature interval. The agreement is quite reasonable. A "stabilized-jellium" density-functional theory (DFT) calculation [20] for Na gives a coefficient of the same magnitude. This affirms that for these alkali metals thermal expansion is the dominant factor behind the temperature-induced shifts in their work functions, and also that for nanoparticle sizes studied here the thermal expansion coefficient is close to that of the bulk material.

**Table II**. Work functions and their temperature coefficients in the solid phase, derived from the data in Fig. 1, and comparison with the image charge model and earlier literature information

|    | Extrapolated $W$ (eV), this work | | $W$ (eV) in reviews [51,52] | $dW/dT$ (eV/K), this work | $dW/dT$ (eV/K), from model [21] | $dW/dT$ (eV/K) in [17,24] |
|----|---------------------|---------------------|-------------|-----------------|-----------------|-----------------|
|    | $T \to 0$ K         | $T \to 293$ K       |             |                 |                 |                 |
| Na | $2.802 \pm 0.009$   | $2.743 \pm 0.016$   | 2.54 – 2.75 | $-1.7\times10^{-4}$ | $-2.0\times10^{-4}$ | $-5.1\times10^{-4}$, $-4.0\times10^{-4}$ |
| K  | $2.390 \pm 0.012$   | $2.317 \pm 0.023$   | 2.29 – 2.30 | $-2.5\times10^{-4}$ | $-2.7\times10^{-4}$ | $-2.6\times10^{-4}$, $-5.3\times10^{-4}$ |

In the molten phase the work function displays a steeper temperature dependence, see Fig. 1. This reflects the greater thermal expansivity of the disordered liquid phase. Because the work function is sensitive to both the electron density and the surface potential barrier—both of which are affected by thermal expansion—this results in a more pronounced temperature dependence in the liquid state. The thermal expansion coefficient of liquid K is higher than that of liquid Na [47,53], hence the slope of its $W(T)$ in this phase is greater. As concerns the magnitudes of these slopes, estimates based either on a transposition of the image charge model [21] to the liquid-state densities [46] or on a variational DFT treatment of a two-component model [20] are within an order of magnitude of the data in Fig. 1 but mostly not in numerical agreement. Therefore a careful theoretical analysis of the liquid metal work functions at different temperatures would be of interest.

*Conclusions*. We have demonstrated that temperature-dependent measurements of the work function of isolated metal nanoparticles can attain the sensitivity and accuracy to resolve both the effect of their thermal expansion and the occurrence of a structural phase transition. The results spotlight the value of the photoelectric effect as a probe of structural changes across phase boundaries and of thermal lattice effects in finite systems. By making use of scaling relations, these features also can be informatively extrapolated to bulk systems.

The experimental data on free Na and K nanoparticles of 7-9 nm diameter reveal discontinuities in the work function as a function of temperature, consistent with melting the onset of melting but occurring significantly below the bulk melting points. The latter is a well-known feature of nanoscale systems, characterized by the Gibbs–Thomson equation.

In a finite-sized nanoparticle the phase transition is broadened, scaling approximately as $\delta T_m \sim 10 T_m/N$, where $T_m$ is the melting point [32]. In the current experiment this implies a transition width of $\delta T_m \sim 0.5$ K. The employed temperature increments of 20 K do not allow this broadening to be observed, but it can become detectable in principle with the present method by using finer temperature steps and adjusting source design to produce smaller particles. This will offer an interesting step in the study of phase change processes on the nanoscale. Furthermore, by optimizing the precision of measurements of the temperature dependence of nanocluster work functions, these can be used to characterize the thermal expansion of quantum-sized systems (see, e.g., [54] for a proof-of-concept experiment).

The use of free nanoparticles also confers benefits that complement conventional surface photoemission studies. As pointed out in the introduction, they derive from the fact that nanoparticle beams provide experimental access to samples that are ultrapure despite being highly chemically reactive. A similar accuracy is challenging to achieve by other means.

For example, by using the photoemission signal to systematically analyze the melting point depression as a function of nanoparticle size in the beam, the Gibbs–Thomson equation can be used, in a direct and non-invasive way, to determine the solid–liquid interfacial energies of strongly reactive elements.

Finally, nanoparticles are a unique resource for exploring photoemission from liquid metals. Bulk liquid metal surfaces are exceptionally reactive, whereas liquid-metal-jet sources have neither been adapted for photoemission spectroscopy (microjets of a strongly metallized solvent are an interesting development in this direction [55]) nor are temperature controlled. These obstacles can be bypassed by employing a beam of metal nanoparticles thermalized to specific temperatures above their melting point. In particular, it would be interesting to explore the work function behavior of bismuth and gallium nanoparticles because in the bulk these materials exhibit anomalous behavior (an increase in density) upon melting [56,57].

**ACKNOWLEDGMENT**

This research was supported by the U.S. National Science Foundation under Grant No. DMR-2003469.

**APPENDIX: IN-FLIGHT EVAPORATIVE COOLING OF NANOPARTICLES.**

The typical time it takes a particle of $N$ atoms, originally at temperature $T$, to evaporate an atom is given by [32]

$$\tau \approx \tau_0 N^{-2/3} e^{D/k_B T}, \tag{8}$$

where the $N^{2/3}$ factor represents the size scaling of the surface area and $D$ is the atom's surface binding energy. For Na this energy and the characteristic time $\tau_0$ have been determined to be 1 eV and $2T^{1/2} \times 10^{-15}$ s, respectively [58]. For a Na particle with $N = 5000$ at $T = 400$ K this translates into an evaporation time of $\tau \approx 0.5$ ms. Therefore during the approximately 10 ms flight time a molten sodium nanoparticle may evaporate on the order of 20 atoms, losing a total of ~20 eV of internal energy. With the heat capacity of liquid Na at 400 K of 31.5 J/mol-K [59], this translates into a temperature decrease of $\approx 10$ K, which is not immaterial but is a relatively small correction at these temperatures. The binding energies [60] and other parameters for potassium near its bulk melting point are not qualitatively different and will lead to a similar conclusion.